\documentclass[12pt]{article}

\usepackage{scicite}
\usepackage{graphicx}

\usepackage{times}

\newenvironment{sciabstract}{%
\begin{quote} \bf}
{\end{quote}}

\newcounter{lastnote}

\title{Smectic to nematic transition of magnetic texture\\
in conical state}

\author
{Masayasu Takeda,$^{1,2\ast}$ Yasuo Endoh,$^{1,3}$ Kazuhisa Kakurai,$^{1}$\\
Yoshinori Onose,$^{4}$ Junichi Suzuki,$^{1,2}$ Yoshinori Tokura$^{4,5}$\\
\\
\normalsize{$^{1}$Quantum Beam Science Directorate, Japan Atomic Energy Agency,}\\
\normalsize{Tokai, Ibaraki 319-1195, Japan}\\
\normalsize{$^{2}$J-PARC Center, Tokai, Ibaraki 319-1195, Japan}\\
\normalsize{$^{3}$International Institute for Advanced Studies, Kizugawa, Kyoto 619-0225, Japan}\\
\normalsize{$^{4}$Department of Applied Physics, University of Tokyo, Tokyo 113-8656, Japan}\\
\normalsize{\& ERATO Multfierroics Project, JST, c/o University of Tokyo, Tokyo 113-8656, Japan}\\
\normalsize{$^{5}$Cross-Correlated Materials Research Group (CMRG), ASI, RIKEN, Wako 351-0198, Japan}
\\
\normalsize{$^\ast$To whom correspondence should be addressed; E-mail:  takeda.masayasu@jaea.go.jp.}
}

\date{}

\begin{document} 


\baselineskip24pt


\maketitle


\begin{sciabstract}
Complex phase transition of magnetic long range order in Fe$_{0.7}$Co$_{0.3}$Si has been investigated by small-angle neutron scattering (SANS) measurements, and found liquid-crystal-like behavior of the magnetic textures for the first time. A variety of the scattering patterns, ring, crescent-shape, diffusive spots etc. appeared by changing temperatures and applying magnetic fields. The change of the scattering pattern can be explained by an analogy of the phase transition of the liquid crystal when the net magnetization of the conical state under the magnetic fields is regarded as the director of the liquid crystals. In addition, the SANS pattern suggests that the mysterious phase just below T$_{\rm C}$  (known as gphase Ah) corresponds to the gSwiss rollh texture which has been discovered in isomorphous compound, FeGe.
\end{sciabstract}

% In setting up this template for *Science* papers, we've used both
% the \section* command and the \paragraph* command for topical
% divisions.  Which you use will of course depend on the type of paper
% you're writing.  Review Articles tend to have displayed headings, for
% which \section* is more appropriate; Research Articles, when they have
% formal topical divisions at all, tend to signal them with bold text
% that runs into the paragraph, for which \paragraph* is the right
% choice.  Either way, use the asterisk (*) modifier, as shown, to

\paragraph*{}
IThe microscopic structure of spin moments can be decisively determined by analyzing the neutron magnetic scattering diffraction patterns, and the diffractometry for helimagnetic and/or antiferromagnetic materials is particularly important since their microscopic structures in such materials are usually complex. Furthermore, unusual bulk properties such as the anomalous Hall effect as well as the quantum phase transition under the high pressure in the helimagnetic phase of MnSi family has currently attracted great interest, which is all attributed to the unique crystal structure in the MnSi family ({\it 1, 2\/}).  Here we present unique diffraction patterns directly reflected by the existence of complicated magnetic texture in the bulk material of Fe$_{1-x}$Co$_{x}$Si, isomorphous of MnSi showing a helimagnetic spiral spin structure ({\it 3, 4\/}). The MnSi family including Fe$_{1-x}$Co$_{x}$Si has cubic crystal structure with low symmetry of  B$_{{2}_{1}3}$ (B20) so that the noncentro-symmetry or the lack of the inversion symmetry of the crystal lattice with a single chiral lattice system ({\it 5\/}) provides the long range spiral spin modulation by the action of Dzyalosinsky-Moriya (DM) interaction even though the fundamental magnetic property is ferromagnetic ({\it 6, 7\/}).

\paragraph*{}
Polarized neutron magnetic scattering determines the chirality of the spiral structure in which MnSi and Fe$_{1-x}$Co$_{x}$Si show opposite chirality; counterlock wise in MnSi and clockwise in FeCoSi indicating the opposite sign of DM term in each crystal possibly due to the sign change of the spin-orbit coupling in each crystal ({\it 8\/}) . Though a number of investigations of the microscopic spin structure in both MnSi family have been performed, none of these studies has paid much attention to the relevance of the magnetic texture. Quite recently, however, Uchida et al. presented the real space image of the helical spin order in Fe$_{0.5}$Co$_{0.5}$Si by the Lorentz microscopy where the electron beams are diffracted by the existence of the local magnetic field in the magnetic specimen, which gives a contrast in intensity of transmitted image corresponding to the magnetization density distribution in the crystal ({\it 9\/}). The long range spiral spin modulation gives rise to the sinusoidal modulation of the net magnetization projected on an axis perpendicular to the modulation direction (a-axis) as a stripe pattern. In addition the dark strings crossing several modulation periods spread in the 2D image and the spin stripes exhibit rather complicated feature by changing temperature indicating that helical magnetic domains thermally fluctuate. 

\paragraph*{}
The Lorenz microscopy showing the magnetic density profile in real space is a very powerful tool but the actual experiment is limited in very low external magnetic field. We were motivated by the work of the Lorenz microscopy on the magnetic texture suggesting that the magnetic domain configuration with a lot of defects like edge dislocations might influence the microscopic spin structure itself which can be detected by the small-angle neutron scattering (SANS). We expect to detect a fingerprint of the magnetic textures in the diffraction pattern, since the size of magnetic domains is nearly the same order of the modulation period of spiral modulation in Fe$_{1-x}$Co$_{x}$Si. 

\paragraph*{}
Prior to the present studies we also looked at the previous neutron scattering studies of Fe$_{1-x}$Co$_{x}$Si ({\it 8, 10\/}) and focused the evolution of the magnetic texture in the spiral phase of Fe$_{0.7}$Co$_{0.3}$Si by controlling temperature and external magnetic field ($T < 50 {\rm K},  \mu_{0}H < 0.2 {\rm Tesla}$). First we present here the SANS patterns with polarized neutrons projected on the 2-dimensional position sensitive detector (2D-PSD) in Fig.\ 1. The scattered intensities coded by different colors show the scattering profile on 2D phase space perpendicular to the incident neutron wave vector, ki. The figure was taken at 10 K, and in zero field ($\mu_{0}H = 0$) (left pair) and at the finite field ($\mu_{0}H  = 0.125 {\rm Tesla}$) with switching the neutron polarization activated by the spin flipper either parallel or antiparallel to the horizontal axis where the crystalline [110] is also parallel to the horizon. The intensity distribution of the pair of the scattering profile is inverted with each other, though the symmetry axis is slightly declined from the exact vertical axis. The fact should be emphasized that the scattered intensities are weak when the scattering vector ($\vec{\kappa}$ ) is along the neutron polarization ($\vec{P}$), and that the scattering intensities disappear when the two vectors are exactly parallel to the magnetic field ($\vec{P} // \vec{\kappa} // \vec{H}$ ). It clearly shows the existence of the chirality of the helix (see Figs. 1C and 1D).  
\begin{figure}[ht]
\centering
\includegraphics[scale=0.35]{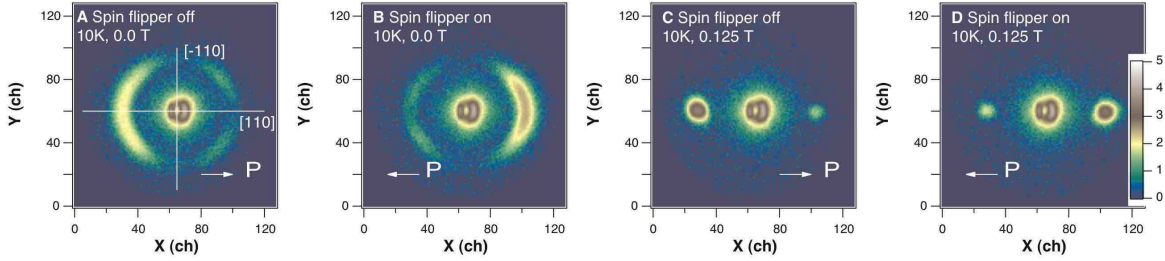}
\caption{Color coded scattering profiles on 2D-PSD for SANS with polarized neutrons at 10 K.
Left set (A, B) show $\mu_{0}H = 0$ after zero field cooling from the ambient temperature, and right set (C, D) in $\mu_{0}H = 0.125$ Tesla. Horizontal axis is nearly parallel to [110] axis of the cubic crystal and vertical axis is exactly parallel to [-110]. Neutron polarization is either parallel (in A and C) or antiparallel (in B and D). Strong intensities very near the center are due to the transmitted neurons uncovered by the attenuator mask.}
\label{fig:1}
\end{figure}

\paragraph*{}
The chirality of the helical spin order in Fe$_{1-x}$Co$_{x}$Si is thus confirmed to be opposite in the helical state of MnSi ({\it 4, 8\/}). In other words, the sign of the chiral order term is reversed in two materials of the same family. The ring or circular scattering shape as shown in Fig. 2 continuously exists up to 40 K at $\mu_{0}H = 0$, though the scattering intensities along the ring are heterogeneous as temperature increases. Above 40 K there appear brighter spots along the crystalline [100]  which indicates the preferred orientation of the modulation wave vector.
\begin{figure}[ht]
\centering
\includegraphics[scale=0.35]{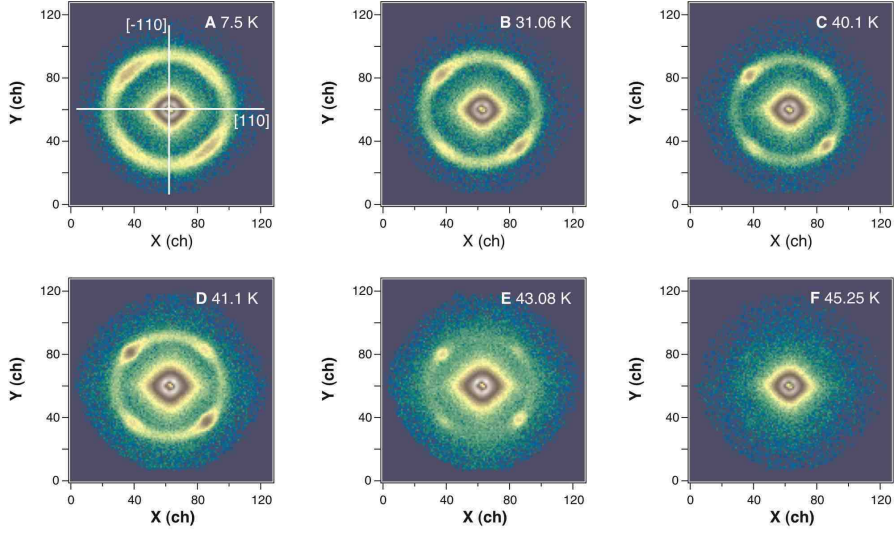}
\caption{Color coded SANS patterns with unpolarized neutrons on the 2D-PSD at various temperatures under the zero external magnetic field ($\mu_{0}H = 0$). A 7.5 K, B 31.06 K, C 40.1 K, D 41.1 K, E 43.08 K and F 45.25 K. The orientation of the crystal is the same as in Fig. 1.}
\label{fig:2}
\end{figure}
It should be emphasized here that the circular scattering profile or the ring-shape at $\mu_{0}H = 0$ after zero-field cooled virgin state was confirmed to be only an intersection on the 2D detector plane of the spherical shell of scattering profile. The ring shape is quite independent of the azimuthal angle for more than 30 degrees around the vertical axis. The comparison of integrated intensities along the circle (shell) with respect to those of the double dots in the higher external field validates the above conjecture; the ring profile is the intersection of the spherical shell of the scattering profile on the detector plane. 

\paragraph*{}
Next the magnetic field hysteresis at the lowest temperature as well as at 40 K is presented in Fig.  3. The representative patterns are illustrated here showing clear change of scattering profiles from circular, ellipsoidal, crescent, and double dots with field application along [100] axis at 4.5 K. The same tendency was also observed in $\mu_{0}H // [110]$. Once the crystal is experienced in the magnetic field, the conical state is stabilized so that it essentially shows no change with decreasing field, giving rise to the distinct hysteresis up to ~ 30 K. The change of the diffraction patterns shows the robust effect that the helical spin structure in the bulk forms the magnetic texture. In particular the diameter of the ring (spherical shell in fact) is shorter in the reciprocal space than the wave vector of the conical spin order. Therefore during the transition from the ring to the double spot by increasing the external fields, an elliptic diffraction shape with shorter vertical and longer horizontal axes emerges. 
\begin{figure}[ht]
\centering
\includegraphics[scale=0.25]{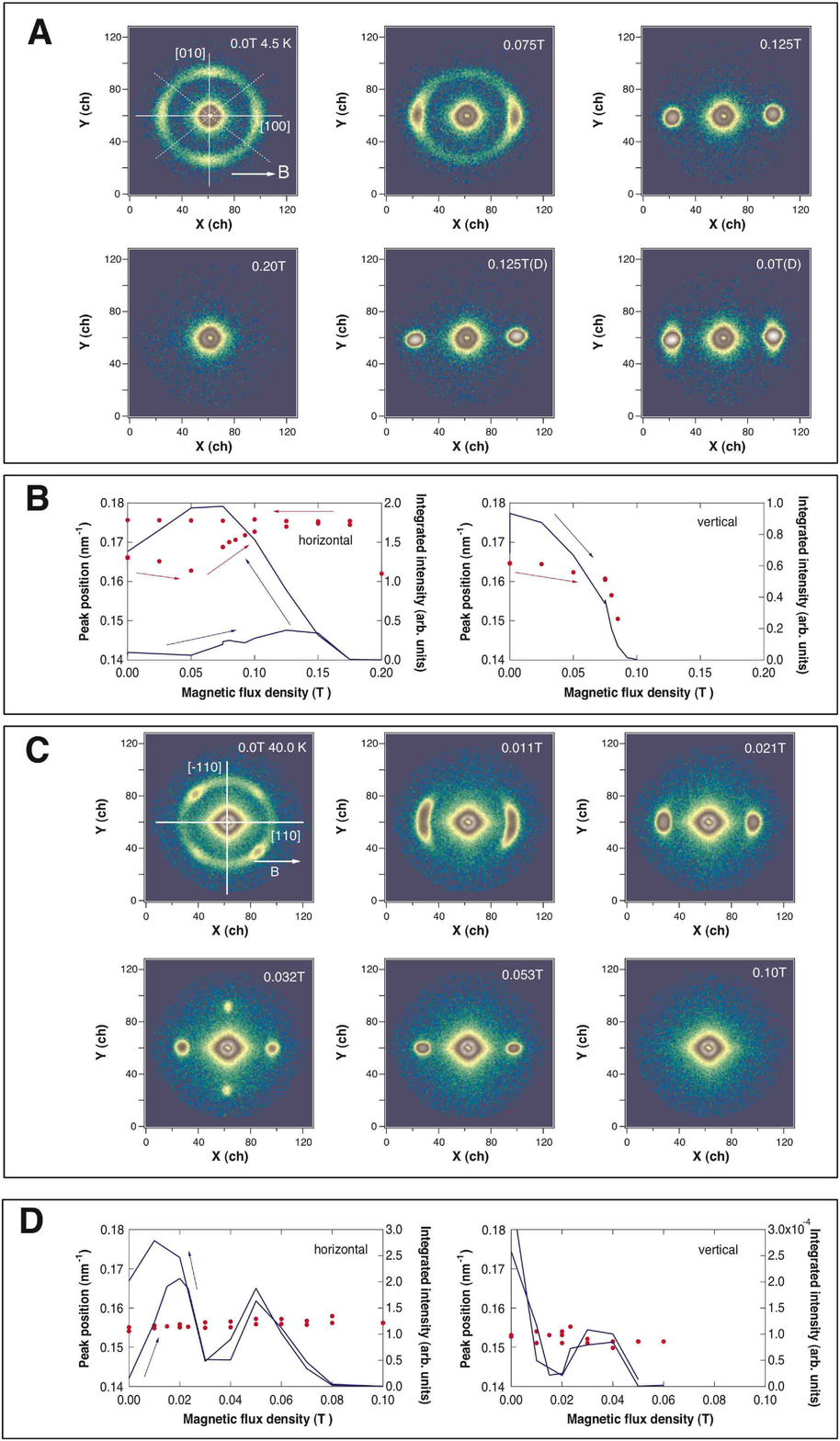}
\caption{Magnetic field dependence of the color coded scattering profiles with unpolarized neutron beam on 2D-PSD, peak positions and integrated intensities. The external magnetic fields were increased from $\mu_{0}H = 0$ to 0.2 Tesla, and then reduced to  $\mu_{0}H = 0$ at 4.5K (A) and at 40 K (C). Scattering intensities (solid lines) and wave vector (solid circles) of the maximum scattering intensity along the parallel and perpendicular directions to horizontal axis (// $\vec{H}$ ) at 4.5 K are plotted in (B) and those at 40 K are plotted in (D) where both horizontal and vertical directions are shown. }
\label{fig:3}
\end{figure}

\paragraph*{}
In contrast, around 40 K the field dependence of the diffuse patterns clearly changes from the lower temperature patterns as shown in Fig. 3C. As increasing the external field above 0.02 T, the crescent like pattern turns out to become distinct 4 diffusive spots, one pair along the field direction and the other pair perpendicular to the field and the hysteresis behavior is not discernible. The similar tendency is common with the change of the field direction, though the measurements were limited for a couple of crystal orientations, along [110] and [100]. This phase is known as gphase Ah of FeCoSi, which was first observed in MnSi ({\it 3\/}). However, intensities of the vertical pair perpendicular to the field direction are more than 5 times weaker than those of the horizontal pair, which suggests that the vertical pair can be the intersection of the circular ring pattern extended perpendicular to the 2D detector plane. Furthermore, the hysteresis behavior could not be observed at this temperature. Combined with the sudden drop in net magnetization in the gphase Ah (~ 40 K, ~ 0.02 T) ({\it 8\/}), the change of the diffraction patterns suggests that the magnetic texture shows the phase transition from 1D stripe to a cylindrical one as defined as gSwiss rollh in FeGe ({\it 11\/}) with the change of the spiral direction from parallel towards perpendicular direction. Schematic representation of each texture in real space is illustrated in Fig. 4.
\begin{figure}[ht]
\centering
\includegraphics[scale=0.2]{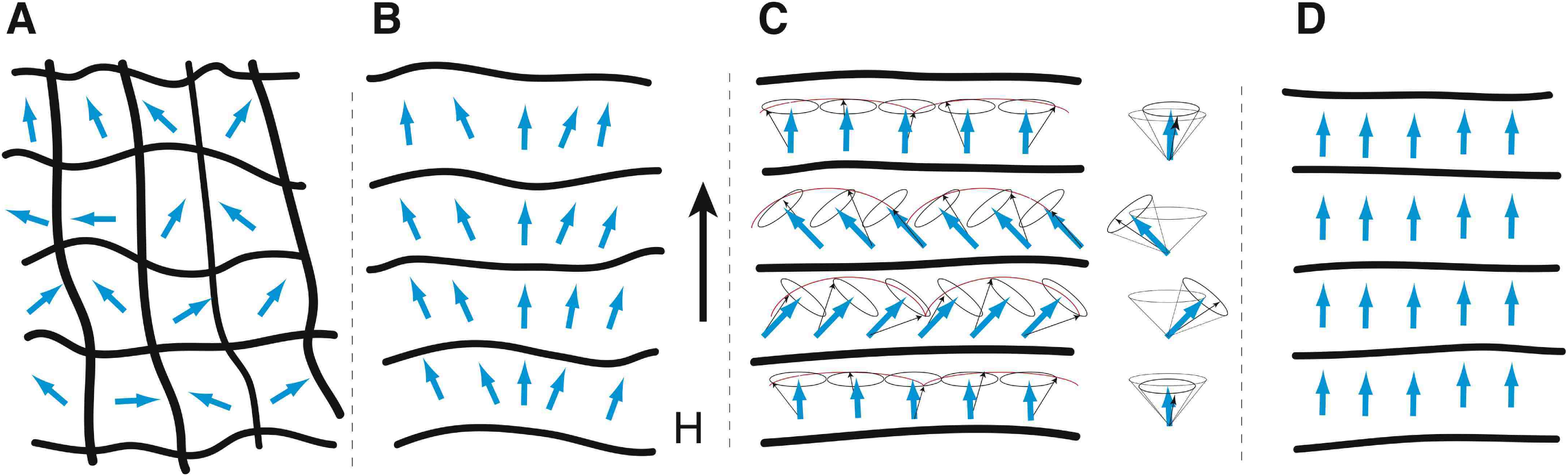}
\caption{2D image of the magnetic textures under typical conditions in real space. A The random pattern, B crescent pattern, C ring pattern, and D diffuse spotty pattern, corresponding to the virgin state at lowest temperature, lower field phase, gphase Ah at~40 K, and phase in the high magnetic fields, respectively. The blue-sky arrows indicate the direction of the propagation vector ($\vec{q}$) of the helical spin order in A and the cone axis of spiral or the ferromagnetic component in a texture in B, C and D. $\vec{q}$ rotates around the field direction in C where the black arrows represent the direction of spins. In addition to the conical order, these spins show a cycloidal ordering perpendicular to the field direction. }
\label{fig:4}
\end{figure}

\paragraph*{}
To conclude, the scattering patterns detected by the SANS from Fe0.7Co0.3Si single crystal probes a fingerprint of versatile magnetic textures. At the virgin state at the low temperature below ~ 30 K and in zero external magnetic field, the magnetic texture is either randomly distributed though {100} direction seems to be preferable or spherically oriented layers like an onion. Furthermore, the unique phase transition of magnetic texture emerges by changing magnetic field as well as temperature, which is very similar to the phase transition of liquid crystal, like smectic and nematic phase transition. 

\paragraph*{}
In the nematic phase of the liquid crystal, the long axis of molecules align in a unique direction without the positional order, while in the smectic phase the molecules form a layer structure as well as the order of the orientation in the nematic phase. In this correspondence, the net magnetization determined as the cone axis of the spiral spin in a magnetic texture can be viewed as the director of the liquid crystal (the unit vector of the preferred orientation). Note that the virgin state in the zero field, the chiral vector perpendicular to the helix plane is regarded as the order parameter, since the net magnetization is zero.

\paragraph*{}
It should be emphasized here that the SANS measurement provides for the first time the change of diffraction patterns by controlling both temperature and the external magnetic field, which is the finger print of the magnetic texture in the spiral spin state of FeCoSi. The fact is that a long period of the helical spin order in FeCoSi as well as MnSi is nearly the same order of the magnetic domain size and that the magnetic diffraction pattern shows the reflection of the magnetic texture. Various issues on the magnetic phase change in the helical magnetic phase (below 45 K, 0.2 T) as well as the unusual bulk properties should be considered in the light of magnetic texture characters presented here. 

\paragraph*{Acknowledgments.}
The authors thank K. Ishimoto for showing his Ph. D thesis on the general magnetic properties of FeCoSi and N. Nagaosa for valuable discussions.\\

{\bf References and Notes}
\begin{enumerate}
\item C. Pfleider, G. J. McMullan, S. R. Jurian, G. G. Lonzarich, {\it Phys.\ Rev.\ B} {\bf 55} 8330 - 8338 (1997).
\item M. Lee, Y. Onose, Y. Tokura, N. P. Ong, {\it Phys.\ Rev.\ B}, {\bf 75}, 172403-1-4  (2007).
\item Y. Ishikawa, K. Tajima, D. Bloch, M. Roth, {\it Sol.\ Stat.\ Commun.} {\bf 19}, 525 - 528 (1976).
\item M. Ishida, Y. Endoh, S. Mitsuda, Y. Ishikawa, M. Tanaka, {\it J.\ Phys.\ Soc.\ Jpn.} {\bf 54}, 2975 - 2982 (1985).
\item M. Tanaka, H. Takayoshi, M. ishida, Y. Endoh, {\it J.\ Phys.\ Soc.\ Jpn.} {\bf 54}, 2970 - 2974 (1985).
\item M. Kataoka, O. Nakanishi, {\it J.\ Phys.\ Soc.\ Jpn.} {\bf 50}, 3888 - 3896 (1981).
\item P. Bak, M. H. Jensen, {\it J.\ Phys.\ C} {\bf 13}, L881 - L885 (1981).
\item K. Ishimoto, thesis, Tohoku University (1995).
\item M. Uchida, Y. Onose, Y. Matsui, Y. Tokura, {\it Science} {\bf 311}, 359 - 361 (2006).
\item K. Ishimoto et al., {\it Physica B} {\bf 213 \& 214}, 381 - 383 (1995).
\item M. Uchida, N. Nagaosa, J. P. He, Y. Kaneko, S. Iguchi, Y. Matsui, Y. Tokura, {\it Phys.\ Rev.\ B} {\bf 77}, 184402-1-5 (2008).
\end{enumerate}

\end{document}